\documentclass[aps,twocolumn,
superscriptaddress,
footinbib,
prb]{revtex4-2}

\newcommand{\cref}[1]{Ref.~\cite{#1}}

\newcommand{\cact}[1]{\hyperref[#1]{#1}}
\newcommand{\ignore}[1]{}

\usepackage{amsmath,amssymb,bm}
\usepackage{graphicx}
\usepackage[usenames,dvipsnames]{color}
\usepackage{hyperref}
\usepackage{natbib}
\usepackage{xcolor}
\usepackage{braket}
\usepackage{tikz}
\usetikzlibrary{shapes}
\usetikzlibrary{calc}

\usepackage{MnSymbol} % for \llangle

\usepackage[utf8]{inputenc} 

\renewcommand{\vec}{\mathbf}

\begin{document}

\title{Time-dependent variational Monte Carlo without bias}

\date{\today}

\author{Wladislaw Krinitsin}
    \affiliation{University of Regensburg, 93053 Regensburg, Germany}
    \affiliation{Forschungszentrum J\"ulich GmbH, Peter Gr\"unberg Institute, Quantum Control (PGI-8), 52425 J\"ulich, Germany}
\author{Markus Schmitt}
\email{markus.schmitt@ur.de}
    \affiliation{University of Regensburg, 93053 Regensburg, Germany}
	\affiliation{Forschungszentrum J\"ulich GmbH, Peter Gr\"unberg Institute, Quantum Control (PGI-8), 52425 J\"ulich, Germany}

\begin{abstract}
When combined with highly expressive ansatz functions such as neural quantum states, variational Monte Carlo (VMC) constitutes a versatile numerical approach to tackle the quantum many-body problem in and out of equilibrium.
However, its traditional formulation exhibits a subtle estimation bias leading to inaccuracies, which can be particularly detrimental when addressing real time dynamics. 
In this work, we investigate two avenues to circumvent said estimation bias. 
First, we propose an unbiased variant of time-dependent VMC using self-normalized importance sampling with respect to a cutoff-based deformation of the Born distribution. 
We demonstrate the feasibility and accuracy of the approach in pathological and generic cases of quench dynamics. 
Furthermore, we explore an alternative sampling strategy based on active learning via the tensor cross interpolation (TCI).
While we find that our choice of tensor network architecture lacks the required low rank property, the proposed TCI-based algorithm complements the conventional importance sampling paradigm, providing an alternative perspective that may be further explored in future work.
\end{abstract}

\maketitle  

\section{Introduction} 
The simulation of strongly correlated quantum many-body systems remains one of the central challenges in condensed matter physics. 
The exponential growth of the Hilbert space with system size limits exact approaches to the smallest systems, forcing the development of sophisticated approximation methods. 

In recent years, the introduction of Neural Quantum States (NQS) -- artificial neural networks used as variational wave functions -- have emerged as a promising new approach for compressed quantum state representations \cite{carleo}. 
Leveraging the universal approximation theorem and the ability to capture complex, non-local correlations \cite{sharir2022neural}, NQS have been shown to achieve state-of-the art accuracy for a wide range of models and physical situations, including strongly correlated spin systems~\cite{hibat2020recurrent,schmitt2020quantum,burau2021unitary,viteritti2023transformer,donatella2023dynamics,chen2024empowering,Chen2025,viteritti2026approaching}, electrons~\cite{RobledoMoreno2022,nys2024ab,gu2025,Chen2025_pfaffian,GauvinNdiaye2025,Viteritti2026}, or lattice gauge theories \cite{Luo2021,lin2024real}. 
Dynamical situations, where established numerical methods face severe challenges due to rapidly growing entanglement or due to the Monte Carlo sign problem, is a particularly relevant field of application for NQS \cite{hartmann2019neural,schmitt2020quantum, reh2021time,luo2022autoregressive,MendesSantos2023, donatella2023dynamics,Medvidovic2023,MendesSantos2024,lin2024real,nys2024ab,Schmitt2025,Chen2025,VandeWalle2025,Sinibaldi2026}.
Typically, these neural networks are optimized using the time-dependent variational principle (TDVP), which allows to perform ground state search or dynamics via imaginary- or real-time evolution within the variational manifold, respectively.

Despite the remarkable successes of NQS, a fundamental and often overlooked technical bottleneck persists: the evaluation of the TDVP step relies on Monte Carlo techniques, which is commonly referred to as time-dependent variational Monte Carlo (t-VMC) \cite{Carleo2012}, to estimate all quantities of interest by sampling from the probability distribution defined by the squared amplitude of the wave function---the Born distribution.
This approach may, however, exhibit a systematic estimation bias \cite{sinibaldi2023unbiasing}, the result of a mismatch between the supports of the Born distribution and the quantity to be estimated.
While this bias will vanish in the thermodynamic limit for typical many-body states \cite{Schmitt2025}, it leads to wrong evaluations of the TDVP update step when encountering pathological states \cite{sinibaldi2023unbiasing}. This is especially detrimental for real-time dynamics where a high level of accuracy needs to be maintained at each step.

In this paper, we investigate different approaches to circumvent said estimation bias, considering quench dynamics in the one dimensional transverse field Ising model (TFIM) as an exemplary application. 
On the one hand, we introduce an alternative, cutoff-based sampling probability through the framework of importance sampling.
Similar approaches have been attempted in the past~\cite{ruiz2016,Inui2021,misery2025,wan2026}: while they mainly focus on decreasing the statistical noise of the estimator, the recently published Ref.~\cite{wan2026} also reduces the estimation bias via post-processing of the samples given a problem specific, sparse kernel. 
In contrast, our work shows that a suitably chosen probability function allows us to directly adapt the support of the sampling distribution in order to obtain an unbiased estimator, while preserving the favorable computational scaling of Monte Carlo methods.
On the other hand, we propose an alternative to the sampling paradigm, using tensor cross interpolation (TCI) \cite{Fernandez2025} as an active learning approach to obtain low-rank approximations of the relevant objects, that can be evaluated efficiently by tensor contraction.
Our empirical study, however, indicates that the relevant objects within the t-VMC framework lack the low-rank property required for the TCI approach.

We start by sketching the general framework and working equations of the TDVP, followed by an introduction of the proposed alternative approaches in the form of importance sampling and TCI in Section \ref{ref:methods}.
The results outlining their efficacy in enabling an unbiased TDVP are presented in Section \ref{sec:results}.
We close this work by providing a summary and discussion of the main findings.

\section{Methods} 
\label{ref:methods}

\subsection{Time Dependent Variational Principle}

In the following, we will be interested in solving the time-dependent Schrödinger equation
\begin{align}
    i\frac{d}{dt}\ket{\psi}=\hat H\ket{\psi}\, .
\end{align}
Approximate solutions within the manifold of states obtained from a variational ansatz $\ket{\psi_\theta}$ can be found by employing a TDVP, in order to obtain the corresponding optimal trajectory in parameter space, $\theta(t)$.
In infinitesimal formulations which assume, that optimal solutions for $\theta(t)$ are continuous in time, the quantum geometric tensor (QGT)
\begin{align}
    \hat{\mathcal{S}}_{k,k'}=\frac{\braket{\partial_k\psi_\theta|\partial_{k'}\psi_\theta}}{\langle\psi_\theta|\psi_\theta\rangle}-\frac{\braket{\partial_k\psi_\theta|\psi_\theta}}{\langle\psi_\theta|\psi_\theta\rangle}\frac{\braket{\psi_\theta|\partial_{k'}\psi_\theta}}{\langle\psi_\theta|\psi_\theta\rangle}\label{eq:Q}
\end{align}
and the force vector
\begin{align}
    \hat{\mathcal{F}}_k=
    \frac{\braket{\partial_k\psi_\theta|\hat H|\psi_\theta}}{\langle\psi_\theta|\psi_\theta\rangle}-
    \frac{\braket{\partial_k\psi_\theta|\psi_\theta}}{\langle\psi_\theta|\psi_\theta\rangle}
    \frac{\braket{\psi_\theta|\hat H|\psi_\theta}}{\langle\psi_\theta|\psi_\theta\rangle}\ . \label{eq:F}
\end{align}
play a central role. Here we define the derivative with respect to the $k$-th variational parameter $\partial_k\equiv\partial/\partial\theta_k$.

Assuming a real-valued parameterization, $\theta\in\mathbb R$, one can derive two versions of the time-dependent variational principle, depending on the starting point \cite{Hackl2020}. Each of them leads to a first-order differential equation prescribing the dynamics of the parameters of the form
\begin{align}
    S_{k,k'}\dot\theta_{k'}=F_k\ , \label{eq:vmc_param_update}
\end{align}
where, doubly occurring indices are summed over, and $\dot\theta\equiv\frac{d}{dt}\theta$.
In the following we will be choosing the stationary action principle as the starting point, corresponding to setting $S_{k,k'}=\text{Im}[\hat{\mathcal{S}}_{k,k'}]$ and $F_k=\text{Im}[-i\hat{\mathcal F}_k]$, which leads to an explicitly energy-conserving solution.

When implementing a TDVP with NQS, a central question is how $\hat{\mathcal{S}}_{k,k'}$ and $\hat{\mathcal F}_k$ can be computed efficiently.
As an example, the first term of the force vector~\eqref{eq:F} takes on the following form:
\begin{equation}
    \frac{\braket{\partial_k\psi_\theta|\hat H|\psi_\theta}}{\langle\psi_\theta|\psi_\theta\rangle}
    =
    \sum_{\vec s}\frac{\braket{\partial_k\psi_\theta|\vec s}\braket{\vec s|\hat H|\psi_\theta}}{\langle\psi_\theta|\psi_\theta\rangle}\, . \label{eq:F_first_term}
\end{equation}
Here, $\{\ket{\vec s}\}$ with $\ket{\vec s}\equiv\ket{s_1}\otimes\ldots\otimes\ket{s_N}$ denotes a computational basis of the $N$-body Hilbert space.
While $\braket{\vec s|\hat H|\psi_\theta}=\sum_{\vec s'}\hat H_{\vec s,\vec s'}\, \psi_\theta(\vec s')$ can be calculated efficiently for sparse Hamiltonians, the sum over all computational basis configurations $\vec s$ becomes exponentially prohibitive with increasing system size. 
One way of addressing this problem is to approximate the full sum via a finite number of samples drawn from a suitable probability distribution, leading to the notion of Monte-Carlo sampling.

\subsection{Self-normalized Importance Sampling}
Monte-Carlo sampling techniques allow to approximate expectation values of the form 
\begin{equation}
    \mu(f)\equiv\mathbb{E}_P[f]=\sum_{\vec s} P(\vec s) f(\vec s)\ , \label{eq:MC_sampling}
\end{equation}
by evaluating the sum on $N_S$ samples drawn according to a given, normalized probability distribution $\vec s \propto P(\vec s)$, resulting in the estimator 
\begin{equation}
    \hat{\mu}_P(f)\equiv\ \frac{1}{N_S} \sum_{\vec s \propto P} f(\vec s). \label{eq:MC_sampling_2}
\end{equation}

Generally, Eq.~\eqref{eq:MC_sampling} can be rewritten as an expectation value with respect to an arbitrary probability distribution $Q(\vec s)=q(\vec s)/\mathcal N_q$
\begin{align}
    \mu (f)&= \mathbb{E}_{P}[f]
    =\sum_{\vec s} P(\vec s) f(\vec s) = \sum_{\vec s} Q(\vec s) \frac{P(\vec s)}{Q(\vec s)}f(\vec s) \nonumber\\
    &= \frac{\mathcal{N}_q}{\mathcal{N}_p} \sum_{\vec s} Q(\vec s) \frac{p(\vec s)}{q(\vec s)}f(\vec s) 
    %=\frac{\mathbb{E}_Q \big[p/q \cdot f\big]}{\mathbb{E}_Q \big[p/q\big]}, \label{eq:self_norm_import}
    =\frac{\mathcal{N}_q}{\mathcal{N}_p} \mathbb{E}_Q \big[p/q \cdot f\big], \label{eq:self_norm_import}
\end{align}
with the weights $q(\vec s)$ and the corresponding normalization constant $\mathcal N_q=\sum_{\vec s}q(\vec s)$. 
Analogously, $P(\vec s)=p(\vec s)/\mathcal N_p$.

One distinguishes between two different cases; on the one hand side, when both normalization constants are known, this procedure is referred to as importance sampling. 
In the other case, the ratio in Eq.~\eqref{eq:self_norm_import} needs to be estimated
\begin{equation}
    \frac{\mathcal{N}_q}{\mathcal{N}_p} = \frac{1}{\mathbb{E}_Q \big[p/q\big]},
\end{equation}
which is referred to as self-normalized importance sampling (SNIS)~\cite{Hesterberg_1995,Owen_2013}. 
While the second option allows for a greater flexibility in the choice of the probability distribution $Q(\vec s)$, care needs to be taken in the estimation of the ratio of normalization constants, which we will discuss further below.

The self-normalized estimator derived from Eq.~\eqref{eq:self_norm_import},
\begin{equation}
    \hat{\mu}_Q(f) = \frac{\frac{1}{N_S}\sum_{\vec s \sim Q}\frac{p(\vec s)}{q(\vec s)}f(\vec s)}{\frac{1}{N_S}\sum_{\vec s \sim Q}\frac{p(\vec s)}{q(\vec s)}}\ ,\label{eq:self_norm_estimator}
\end{equation}
constitutes an asymptotically unbiased estimator when the following two conditions are fulfilled: (i) $q(\vec s)>0$ $\forall \vec s$ where the summand $p(\vec s)f(\vec s)$ is non-vanishing and additionally (ii) $q(\vec s)>0$ $\forall \vec s$ where $p(\vec s) > 0$.
The choice of $q(\vec s)$ affects the variance of the resulting Monte-Carlo estimators -- the theoretically optimal variance reduction is achieved by choosing $q(\vec s) \propto p(\vec s)|f(\vec s)-\mu(f)|$, i.e. proportional to the absolute deviation of the summand from its mean. 
Following that prescription exactly is typically unfeasible, since it requires knowing the value of the mean a-priori; however, it can be used as a guiding principle for finding better, variance reducing distributions.
In particular, the dominant contribution to the variance of the sampled estimator~\eqref{eq:self_norm_estimator} can be extracted using the Delta method~\cite{Casella_2024}
\begin{align}
    \mathrm{Var}[\hat{\mu}_Q(f)] &= \frac{1}{N_S} \hat{\sigma}_q^2(f)\label{eq:snis_variance} \\ 
    &= \frac{1}{N_S} \mathbb{E}_Q \Bigg[\left(\frac{p(x)}{q(x)}\right)^2 \left(\frac{f(x)-\mu}{\mathbb{E}_Q[p(x)/q(x)]}\right)^2 \Bigg]\, , \nonumber 
\end{align} 
see~\cite{gravina2025thesis} for a detailed derivation.

\subsection{Variational Monte Carlo without estimation bias}
The idea of VMC approaches to deal with the intractable sum of Eq.~\eqref{eq:F_first_term} is estimating it via stochastic sampling.
As a starting point one may view the expression as an expectation value with respect to the uniform distribution $U(\vec s)=\mathcal N_u^{-1}$ with $\mathcal N_u=2^N$,
\begin{align}
    \frac{\braket{\partial_k\psi_\theta|\hat H|\psi_\theta}}{\langle\psi_\theta|\psi_\theta\rangle}
    % &=\mathcal N_u
    % \sum_{\vec s}U(\vec s)\frac{\braket{\partial_k\psi_\theta|\vec s}\braket{\vec s|\hat H|\psi_\theta}}{\langle\psi_\theta|\psi_\theta\rangle}
    % \nonumber\\
    &=\mathcal N_u\mathbb E_U \left[\frac{\braket{\partial_k\psi_\theta|\vec s}\braket{\vec s|\hat H|\psi_\theta}}{\langle\psi_\theta|\psi_\theta\rangle}\right]
    \ .
    \label{eq:uniform}
\end{align}
Clearly, a stochastic sampling of the uniform distribution in order to estimate the expectation value will likely turn out inefficient due to high variance of the estimator.
Moreover, the normalization $\braket{\psi_\theta|\psi_\theta}$ is often intractable.
Instead, VMC employs an importance sampling approach. 
The expression in Eq.~\eqref{eq:uniform} suggests the Born distribution $P(\vec s)=\frac{|\psi_\theta(\vec s)|^2}{\braket{\psi_\theta|\psi_\theta}}$ as a natural choice for the sampling distribution, yielding
\begin{align}
\left[\frac{\braket{\partial_k\psi_\theta|\hat H|\psi_\theta}}{\langle\psi_\theta|\psi_\theta\rangle}
    \right]_{\mathrm{VMC}}
    &=\mathbb E_P \left[\frac{\braket{\partial_k\psi_\theta|\vec s}\braket{\vec s|\hat H|\psi_\theta}}{|\psi_\theta(\vec s)|^2}\right]
    \nonumber\\
        &=
    \mathbb{E}_{P}
    \Big[\partial_k\left(\log\psi_\theta^*\right)\, \mathcal{E}_{\text{loc}}\Big]\ , \label{eq:F_first_term_impsamp}
\end{align}
where we introduced the local energy $\mathcal{E}_{\text{loc}}(\vec s)=\sum_{\vec s'}\braket{\vec s|\hat H|\vec s'}\frac{\psi_\theta(\vec s')}{\psi_\theta(\vec s)}$ and the derivative of the log wavefunction for convenience.
The natural character of the Born distribution $P(\vec s)$ as sampling distribution lies in the fact, that all normalization constants cancel and the observable appears correlated with $\psi_\theta(\vec s)$---suggesting favorable variance properties.
While this prescription allows for an efficient and elegant evaluation of all relevant expectation values, it has been pointed out recently that these standard VMC estimators are biased if the wave function exhibits roots, i.e., $\psi_{\vec\theta}(\vec s)=0$ for some $\vec s$ \cite{sinibaldi2023unbiasing}.
In such cases, the roots have to be treated separately, such that
\begin{alignat}{2}
    \frac{\braket{\partial_k\psi_\theta|\hat H|\psi_\theta}}{\langle\psi_\theta|\psi_\theta\rangle}
    &=
    \sum_{\vec s\in\{\vec s'| \psi_{\vec\theta}(\vec s')\neq0\}}
    P(\vec s)
    &&\frac{\braket{\partial_k\psi_\theta|\vec s}\braket{\vec s|\hat H|\psi_\theta}}{|\psi_{\vec\theta}(\vec s)|^2}
    \nonumber\\
    &+
    \sum_{\vec s\in\{\vec s'| \psi_{\vec\theta}(\vec s')=0\}}
    &&\frac{\braket{\partial_k\psi_\theta|\vec s}\braket{\vec s|\hat H|\psi_\theta}}{\langle\psi_\theta|\psi_\theta\rangle}
\end{alignat}
In this expression, the former contribution is captured by the standard VMC estimator and the latter constitutes the bias.
This bias appears analogously in the estimator for the QGT and the resulting biased estimation can be highly detrimental for real-time dynamics.
To address this issue, we revisit the ad-hoc choice of the Born distribution for importance sampling and propose a variant that avoids the estimation bias by construction.

Using Eq.~\eqref{eq:self_norm_import} and a distribution $Q(\vec s)$ with support on the whole domain, we can rewrite the expectation value in Eq.~\eqref{eq:F_first_term_impsamp} as
\begin{align}
    \frac{\braket{\partial_k\psi_\theta|\hat H|\psi_\theta}}{\langle\psi_\theta|\psi_\theta\rangle}
    % &=
    % \sum_{\vec s}\frac{\braket{\partial_k\psi_\theta|\vec s}\braket{\vec s|\hat H|\psi_\theta}}{\langle\psi_\theta|\psi_\theta\rangle}
    % \nonumber\\
    % &=
    % \sum_{\vec s}Q(\vec s)\frac{\mathcal{N}_q}{\mathcal{N}_p}\frac{\braket{\partial_k\psi_\theta|\vec s}\braket{\vec s|\hat H|\psi_\theta}}{q(\vec s)}
    % \nonumber\\
    &=
    \mathbb{E}_{Q}
    \Bigg[\frac{\mathcal{N}_q}{\mathcal{N}_p}\frac{\partial_k\psi_\theta^*\, E_\mathrm{loc}}{q}\Bigg]\ , 
\label{eq:F_first_term_snis}
\end{align} 
where $E_{\text{loc}}=\sum_{\vec s'}\braket{\vec s|\hat H|\vec s'}\psi_\theta(\vec s')$.
The explicit form of all other relevant quantities that appear in the TDVP equation can be estimated in the same way as summarized in Appendix \ref{app:derivation_tdvp}.

Note once again the explicit appearance of the ratio of normalization constants $\frac{\mathcal{N}_q}{\mathcal N_p}$, which can in principle be estimated via
\begin{align}
    \frac{\mathcal{N}_q}{\mathcal N_p}
    =\mathbb{E}_Q\Bigg[\frac{p(\vec s)}{q(\vec s)}\Bigg]^{-1}
    \approx \Bigg[\frac{1}{N_S}\sum_{\vec s \sim Q}\frac{p(\vec s)}{q(\vec s)}\Bigg]^{-1}.
    \label{eq:norm_ratio_estimation}
\end{align}
However in generic cases, a reliable estimation will require exponentially many samples, or otherwise lead to large variances and thus statistical fluctuations of the estimator~\eqref{eq:snis_variance}. 
Nonetheless, there are at least three different routes to treat the ratio of normalizations efficiently: (i) Choose $q(\vec s)$ such that $Q(\vec s)\approx P(\vec s)$ $\forall\vec s$, (ii) work with normalized probability distributions such that $\mathcal N_p=\mathcal N_q=1$, or (iii) keep track of the change of the normalization over time as described, e.g., in Refs.~\cite{carleo_phd,MendesSantos2023}.
Option (ii) is facilitated in the context of NQS by autoregressive architectures~\cite{Wu_2019,hibat2020recurrent}.
In the following we will focus on the first route and use it as a guiding principle to find better behaved probability distributions.

The bias issue arises due to the presence of roots in the wavefunction, resulting in a mismatch between the support of the sampling distribution and the summand. 
Ideally, one would like to directly use the structure of the summand as an informed choice for the probability distribution $Q(\vec s)$, which however, is typically not feasible due to it being a complicated and non-trivial function of the spin variables. 

In view of the primary objective of obtaining an unbiased estimator and with the aforementioned condition that $Q(\vec s)\approx P(\vec s)$ in mind, we propose a simpler solution: a slight deformation of the Born distribution ensuring that the probability globally obeys a non-vanishing lower bound $\epsilon>0$.
Thereby, $q(\vec s)$ has support on the whole configuration space.
Concretely,
\begin{align}
    q_\epsilon(\vec s) = 
        \begin{cases}
        |\psi_\theta(\vec s)|^2 & \text{if } \, \frac{|\psi_\theta(\vec s)|^2}{\max_{\vec s} |\psi_\theta(\vec s)|^2} > \epsilon \\
        \epsilon & \text{else}.
    \end{cases} \label{eq:cutoff_prob}
\end{align}
Notice that this choice, depending on $\epsilon$, interpolates between the traditional approach based on the Born distribution for $\epsilon=0$, and uniform sampling for $\epsilon>1$, where each state has the same sample probability.

Since we are typically working with non-normalized wave-functions, the cutoff in Eq.~\eqref{eq:cutoff_prob} is taken with respect to the magnitude of the largest wave-function coefficient as the reference. 
Once again, following this prescription exactly requires knowledge about the full state vector, which we circumvent by setting $\max_{\vec s} |\psi_{\theta}(\vec s)|$ to the magnitude of the largest wave-function coefficient observed in the samples of the previous timestep. 
The validity of that approximation relies on the wave function coefficients typically varying slowly compared to the timesteps of algorithm.

By choosing the cutoff $\epsilon$ sufficiently small, the resulting ratio of normalization constants will typically be close to $\frac{\mathcal{N}_q}{\mathcal N_p}\approx1$, which can be reliably estimated with a limited number of samples.
In practice, we found that the impact of the ratio of normalization factors on the time evolution is additionally suppressed by the structure of the TDVP equation~\eqref{eq:vmc_param_update}, even for larger $\epsilon$.
This can be argued for heuristically, by first noticing that both Eqs.~\eqref{eq:Q} and ~\eqref{eq:F} can be written as covariances, i.e. take on the form
\begin{align}
    \mathbb{E}_P\big[ X Y \big] &- \mathbb{E}_P\big[ X \big]\mathbb{E}_P\big[ Y \big]\nonumber \\
    = \frac{\mathcal N_q}{\mathcal N_p}\Bigg(\mathbb{E}_Q\left[\frac{p}{q} X Y \right] &- \frac{\mathcal N_q}{\mathcal N_p}\mathbb{E}_Q\left[ \frac{p}{q}X \right]\mathbb{E}_Q\left[\frac{p}{q} Y \right]\Bigg)
\end{align}
with accordingly defined variables $X$ and $Y$. The second line introduces the SNIS estimators. The leading prefactor, the ratio of normalization constants, appears in both $\hat{\mathcal{S}}$ and $\hat{\mathcal{F}}$, and as such will cancel in the calculation of the update~\eqref{eq:vmc_param_update}. 
The remaining ratios $\frac{\mathcal N_q}{\mathcal N_p}$ inside the covariances have been numerically observed to be sub-dominant contributions to the TDVP equation, and do not strongly impact the quality of the time evolution.

Having established self-normalized importance sampling in combination with the introduced cutoff-based probability distributions~\eqref{eq:cutoff_prob} as a potential fix for estimation bias, we turn towards exploring an alternative: the TCI, that samples configurations in a targeted manner via active learning.

\subsection{Tensor Cross Interpolation} \label{sec:TCI_intro}

Tensor cross interpolation~\cite{Fernandez2025} (TCI) is a method used to systematically refine a low-dimensional tensor network approximation of any function $f: V \rightarrow W$. 
In contrast to traditional MC sampling, where samples are chosen based on some probability distribution, TCI is typically referred to as an active learning approach --- samples are actively requested and added in an iterative fashion in order to improve the approximation of the target function, constituting an inherently unbiased approach. 

In the following, the TCI of a function $f$ will be denoted by $\mathrm{TCI}_f$. 
Such an approximation is found by updating the tensor network $\mathrm{TCI}_f$ through the iterative addition of so-called pivots -- points $x \in V, y \in W$ where the difference $y \equiv f(x) - \mathrm{TCI}_f(x)$ between the exact function value and the one provided by the tensor network is maximal. The incorporation of pivots is typically done through local updates in a sweep-like fashion, usually increasing the bond dimension of the network.
At these pivots, the tensor network agrees exactly with the original function, while for all the other points in the domain it provides an interpolation.
The algorithm continues adding pivots until a specified bond dimension $\chi_{\max}$ or tolerance $\epsilon_\mathrm{tci}$ is reached, s.t. $ || f- \mathrm{TCI}_f ||_\infty<\epsilon_\mathrm{tci} $.

Using TCI, we can build a matrix product state representation of the functions
\begin{equation}
    f_{E_\mathrm{loc}}(\vec{s}) = \frac{E_\mathrm{loc}(\vec s) - \psi_\theta(\vec s)\, \braket{\psi_\theta|\hat{H}|\psi_\theta}}{\sqrt{\braket{\psi_\theta|\psi_\theta}}} \label{eq:TCI_eps} 
\end{equation}
and
\begin{equation}
    f_\mathrm{grad}(\vec{s}, k) = \frac{\partial_k \psi_\theta(\vec s) - \psi_\theta(\vec s)\, \braket{\psi_\theta|\partial_k\psi_\theta}}{\sqrt{\braket{\psi_\theta|\psi_\theta}}} , \label{eq:TCI_gamma}
\end{equation}
which can subsequently be used to build the QGT $\hat{\mathcal{S}}_{k,k'}$ and force vector $\hat{\mathcal{F}}_k$, as detailed later in this section.

In contrast to the traditional VMC-approach, the norm of the wavefunction needs to be accounted for explicitly during the construction of the TCI approximation. As detailed previously, there are once again various approaches to deal with this issue, e.g. by using autoregressive networks or by keeping track of the change of the normalization of the network over the course of the time evolution.
\begin{figure}[t!]
  \centering
  \begin{tikzpicture}[baseline=-2, scale=0.6, transform shape]
    \node at (-2.5,-0.) {\LARGE $\mathrm{TCI}_{E_\mathrm{loc}}(\vec s) = $};
    \node (e1) [regular polygon, regular polygon sides=4, shape border rotate=270, fill, minimum size=1.8cm] at (0,0){};
    \node (e2) [regular polygon, regular polygon sides=4, shape border rotate=270, fill, minimum size=1.8cm] at (2,0){};
    \node (e3) [regular polygon, regular polygon sides=4, shape border rotate=270, fill, minimum size=1.8cm] at (5,0){};
    \node at (0,0){\Large \color{white} $E_1$};
    \node at (2,0){\Large \color{white} $E_2$};
    \node at (5,0){\Large \color{white} $E_N$};
    
    \node at (-2.8,-3.) {\LARGE $\mathrm{TCI}_\mathrm{grad}(\vec s, k) = $};
    \node (g1) [regular polygon, regular polygon sides=3, shape border rotate=270, draw, minimum size=1.8cm] at (0,-3){};
    \node (g2) [regular polygon, regular polygon sides=3, shape border rotate=270, draw, minimum size=1.8cm] at (2,-3){};
    \node (g3) [regular polygon, regular polygon sides=3, shape border rotate=270, draw, minimum size=1.8cm] at (5,-3){};
    \node (g4) [regular polygon, regular polygon sides=4, shape border rotate=270, draw, minimum size=1.8cm] at (7,-3){};
    \node at (0,-3){\Large $U_1$};
    \node at (2,-3){\Large $U_2$};
    \node at (5,-3){\Large $U_N$};
    \node at (7,-3){\Large $\Gamma_{N+1}$};

    \node at ($(e2)!.5!(e3)$) {\ldots};
    \node at ($(g2)!.5!(g3)$) {\ldots};

    \draw [thick] (e1) -- (e2);
    \draw [thick] (e2) -- +(1.1,0);
    \draw [thick] (e3) -- +(-1.,0);
    \draw [thick] (e1) -- +(0,-1.) node [below]{\LARGE $s_1$};
    \draw [thick] (e2) -- +(0,-1.) node [below]{\LARGE $s_2$};
    \draw [thick] (e3) -- +(0,-1.) node [below]{\LARGE $s_N$};
 
    \draw [thick] (g1) -- (g2);
    \draw [thick] (g2) -- +(1.1,0);
    \draw [thick] (g3) -- +(-1.,0);
    \draw [thick] (g3) -- (g4);
    \draw [thick] (g1) -- +(0,-1.) node [below]{\LARGE $s_1$}; % node [below]{\scriptsize{$\sigma'_{i-2}$}};
    \draw [thick] (g2) -- +(0,-1.) node [below]{\LARGE $s_2$};
    \draw [thick] (g3) -- +(0,-1.) node [below]{\LARGE $s_N$}; % node [below]{\scriptsize{$\sigma'_{i}$}};
    \draw [thick] (g4) -- +(0,-1.) node [below]{\LARGE $k$}; % node [below]{\scriptsize{$\sigma'_{i}$}};
  \end{tikzpicture}
  \caption{MPS representation of $f_{E_\mathrm{loc}}$ and $f_\mathrm{grad}$. The orthogonality center of the latter is moved to the site carrying the neural network parameter index $k$. The resulting isometries are denoted by triangles, general tensors by squares.} \label{fig:mps_eps_gamma} 
\end{figure}
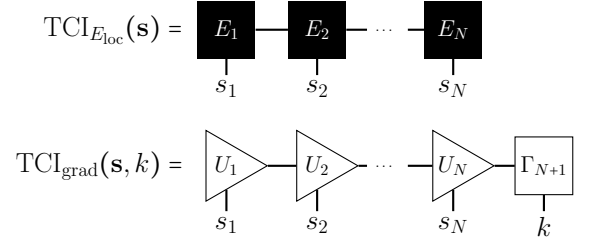
As the purpose of this work is demonstrating the general idea of the method, we choose to calculate the wavefunction norm exactly via the full sum, which is feasible for the system sizes we consider here.
Similarly, the expectation values appearing in Eqs.~\eqref{eq:TCI_eps} and~\eqref{eq:TCI_gamma} are calculated exactly.

In the following we adopt a Matrix Product State (MPS) structure for the TCI representation of $f_{E_\mathrm{loc}}(\vec{s})$ and $f_\mathrm{grad}(\vec{s}, k)$, resulting in one dimensional tensor trains with physical indices consisting of the local spins $\vec s$ and an additional leg for the parameter index $k$ in the case of $\mathrm{TCI}_\mathrm{grad}$; see Fig.~\ref{fig:mps_eps_gamma} for graphical representations. 
Note that this does not encode a low-rank representation of the wave function itself, but for its derivative with respect to the variational parameters of the neural network. It is a priori not clear whether this derivative has such a low rank structure, and will be discussed later in the results section. 

We furthermore want to emphasize that the network architecture chosen for the TCI may drastically impact the convergence properties of the approximation. While the proposed MPS structure is the easiest to implement from an algorithmic perspective, the inherent asymmetry between the the indices for the spins and the variational parameter leads to a suboptimal distribution of the bond dimensions, which can potentially be circumvented by employing a tailored, e.g. tree-like or network topology to be explored in future work.

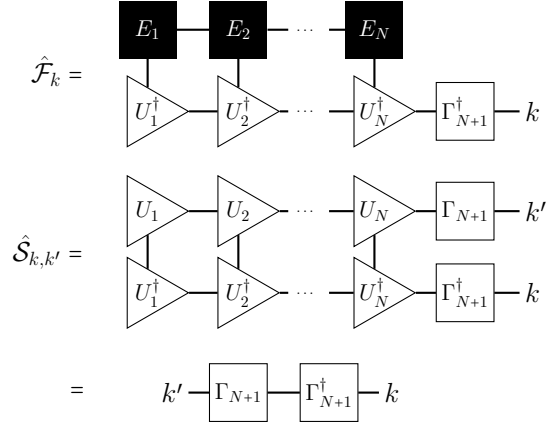
\begin{figure}[t!]
  \centering
  \begin{tikzpicture}[baseline=-2, scale=0.6, transform shape]
    \node at (-2.0,-0.9) {\LARGE $\hat{\mathcal{F}}_k = $};
    \node (fe1) [regular polygon, regular polygon sides=4, shape border rotate=270, fill, minimum size=1.8cm] at (0,0){};
    \node (fe2) [regular polygon, regular polygon sides=4, shape border rotate=270, fill, minimum size=1.8cm] at (2,0){};
    \node (fe3) [regular polygon, regular polygon sides=4, shape border rotate=270, fill, minimum size=1.8cm] at (5,0){};
    \node at (0,0){\Large \color{white} $E_1$};
    \node at (2,0){\Large \color{white} $E_2$};
    \node at (5,0){\Large \color{white} $E_N$};
    
    \node (fg1) [regular polygon, regular polygon sides=3, shape border rotate=270, draw, minimum size=1.8cm] at (0,-1.8){};
    \node (fg2) [regular polygon, regular polygon sides=3, shape border rotate=270, draw, minimum size=1.8cm] at (2,-1.8){};
    \node (fg3) [regular polygon, regular polygon sides=3, shape border rotate=270, draw, minimum size=1.8cm] at (5,-1.8){};
    \node (fg4) [regular polygon, regular polygon sides=4, shape border rotate=270, draw, minimum size=1.8cm] at (7,-1.8){};
    \node at (0,-1.8){\Large $U_1^\dag$};
    \node at (2,-1.8){\Large $U_2^\dag$};
    \node at (5,-1.8){\Large $U_N^\dag$};
    \node at (7,-1.8){\Large $\Gamma_{N+1}^\dag$};
    
    \node at (-2.2,-4.9) {\LARGE $\hat{\mathcal{S}}_{k,k'} = $};
    \node (sg1) [regular polygon, regular polygon sides=3, shape border rotate=270, draw, minimum size=1.8cm] at (0,-4){};
    \node (sg2) [regular polygon, regular polygon sides=3, shape border rotate=270, draw, minimum size=1.8cm] at (2,-4){};
    \node (sg3) [regular polygon, regular polygon sides=3, shape border rotate=270, draw, minimum size=1.8cm] at (5,-4){};
    \node (sg4) [regular polygon, regular polygon sides=4, shape border rotate=270, draw, minimum size=1.8cm] at (7,-4){};
    \node at (0,-4){\Large $U_1$};
    \node at (2,-4){\Large $U_2$};
    \node at (5,-4){\Large $U_N$};
    \node at (7,-4){\Large $\Gamma_{N+1}$};

    \node (s2g1) [regular polygon, regular polygon sides=3, shape border rotate=270, draw, minimum size=1.8cm] at (0,-5.8){};
    \node (s2g2) [regular polygon, regular polygon sides=3, shape border rotate=270, draw, minimum size=1.8cm] at (2,-5.8){};
    \node (s2g3) [regular polygon, regular polygon sides=3, shape border rotate=270, draw, minimum size=1.8cm] at (5,-5.8){};
    \node (s2g4) [regular polygon, regular polygon sides=4, shape border rotate=270, draw, minimum size=1.8cm] at (7,-5.8){};
    \node at (0,-5.8){\Large $U_1^\dag$};
    \node at (2,-5.8){\Large $U_2^\dag$};
    \node at (5,-5.8){\Large $U_N^\dag$};
    \node at (7,-5.8){\Large $\Gamma_{N+1}^\dag$};
    
    \node at (-1.55,-8) {\LARGE $ = $};
    \node (s3g4) [regular polygon, regular polygon sides=4, shape border rotate=270, draw, minimum size=1.8cm] at (2,-8){};
    \node at (2,-8){\Large $\Gamma_{N+1}$};

    \node (s4g4) [regular polygon, regular polygon sides=4, shape border rotate=270, draw, minimum size=1.8cm] at (4,-8){};
    \node at (4,-8){\Large $\Gamma_{N+1}^\dag$};
    
    \node at ($(fe2)!.5!(fe3)$) {\ldots};
    \node at ($(fg2)!.5!(fg3)$) {\ldots};
    % \node at (5.35,-0.5) {\ldots};
    \node at ($(sg2)!.5!(sg3)$) {\ldots};
    \node at ($(s2g2)!.5!(s2g3)$) {\ldots};

    \draw [thick] (fe1) -- (fe2);
    \draw [thick] (fe2) -- +(1.1,0);
    \draw [thick] (fe3) -- +(-1.,0);
    
    \draw [thick] (fg1) -- (fg2);
    \draw [thick] (fg2) -- +(1.1,0);
    \draw [thick] (fg3) -- +(-1.,0);
    \draw [thick] (fg3) -- (fg4);
    
    \draw [thick] (fg4) -- +(1.2,0) node [right]{\LARGE $k$};
    
    \draw [thick] (fe1) -- (fg1);
    \draw [thick] (fe2) -- (fg2);
    \draw [thick] (fe3) -- (fg3);
 
    \draw [thick] (sg1) -- (sg2);
    \draw [thick] (sg2) -- +(1.1,0);
    \draw [thick] (sg3) -- +(-1.,0);
    \draw [thick] (sg3) -- (sg4);
    
    \draw [thick] (sg4) -- +(1.2,0) node [right]{\LARGE $k'$};
    
    \draw [thick] (s2g1) -- (s2g2);
    \draw [thick] (s2g2) -- +(1.1,0);
    \draw [thick] (s2g3) -- +(-1.,0);
    \draw [thick] (s2g3) -- (s2g4);
    
    \draw [thick] (s2g4) -- +(1.2,0) node [right]{\LARGE $k$};

    \draw [thick] (sg1) -- (s2g1);
    \draw [thick] (sg2) -- (s2g2);
    \draw [thick] (sg3) -- (s2g3);
    
    \draw [thick] (s3g4) -- (s4g4);
    \draw [thick] (s3g4) -- +(-1.1,0) node [left]{\LARGE $k'$};
    \draw [thick] (s4g4) -- +(1.1,0) node [right]{\LARGE $k$};
  \end{tikzpicture}
  \caption{MPS contractions between $\mathrm{TCI}_{E_\mathrm{loc}}$ and $\mathrm{TCI}_\mathrm{grad}$, yielding the force vector and QGT, respectively. The calculation of the latter is simplified due to the aforementioned choice of the orthogonality center: by setting it to node carrying the parameter index $k$, the remaining isometries collapse into an identity.} \label{fig:mps_S_F} 
\end{figure}

The covariances~\eqref{eq:Q} and~\eqref{eq:F} are calculated by successively contracting tensors using the zip-up algorithm, allowing for an efficient calculation of sums over the full computational basis in the case of limited bond dimension (see Fig.~\ref{fig:mps_S_F}).

\section{Results} 
\label{sec:results}

In the following we investigate how the proposed methods---cutoff-based importance sampling and TCI---avoid the estimation bias and their effect on the quality of the time evolution of quantum systems. 

\subsection{Importance Sampling}

We start by considering the time evolution of a single spin state $\ket{\psi_{\alpha,\beta}(t)} = \alpha(t) \ket{\downarrow} + \beta(t) \ket{\uparrow}$ under the Hamiltonian $H=\sigma^y$, which has been introduced and studied in~\cite{sinibaldi2023unbiasing} as a pathological example, where the estimation bias has severe effects on the simulation of the dynamics. 
Starting from $\ket{\psi_{\alpha,\beta}(t=0)}=\ket{+}$, the state evolves to $\ket{\psi_{\alpha,\beta}(t=\pi/4)}=\ket{\downarrow}$, under the Rabi oscillation. 
At that point in time, the only non-zero components of the $S$-matrix and force vector originate from the gradient with respect to the $\beta$-parameter. 
Because of that, MC-sampling based on the Born distribution fails to estimate these contributions due to the corresponding wave-function coefficient vanishing, resulting in the dynamics becoming stuck.  

\begin{figure}[t!]
    \includegraphics[width=\columnwidth]{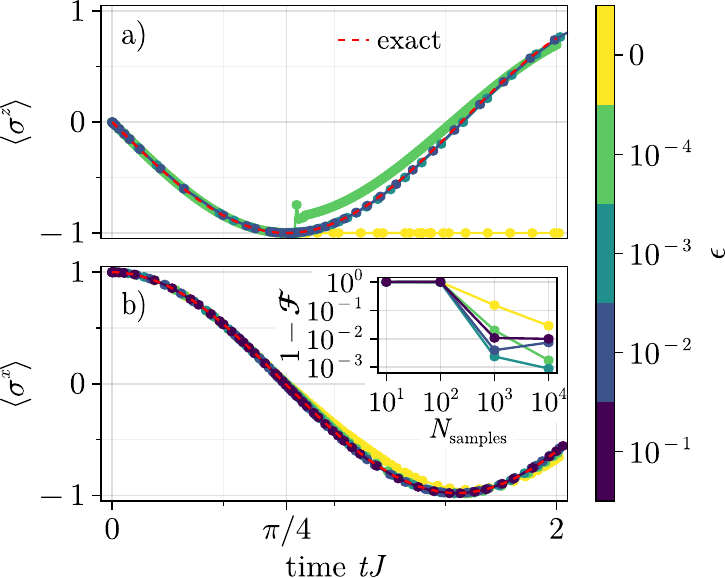}
    \caption{(a) Time evolution of the magnetization for the single spin system rotating around the y-axis. As it approaches the pathological time point $tJ = \pi/4$, sampling based on the Born distribution, denoted by $\epsilon=0$, fails to reproduce the correct dynamics. Choosing a suited cutoff $\epsilon$ fixes the issue. (b) Time evolution of the total x-magnetization for a many-body system of 10 spins evolving under the Hamiltonian~\eqref{eq:bias_interaction} with $J/g=0.1$. The TDVP equation is evaluated using importance sampling for different values of $\epsilon$ and $N_S=10^3$ samples. Choosing $\epsilon=0$ results in noticeable deviations from the reference result. The inset shows infidelities with respect to the exact state after evolving the system until $tJ=2$ for different values of $N_S$. We observe that choosing $\epsilon = 10^{-3}$ improves the infidelity by almost two orders of magnitude for $10^3$ samples compared to the baseline $\epsilon=0$. As such, the unbiased scheme allows to reduce the number of samples compared to Born sampling, without any loss in accuracy.} \label{fig:bias_cutoff}
\end{figure}

This behavior, as well as the impact of importance sampling based on the cutoff-based probability distribution for different $\epsilon$ is shown in Fig.~\ref{fig:bias_cutoff}~a): by introducing a finite cutoff parameter, the sampler is able to pick up the relevant gradient components and recover the correct dynamics.

While this is an illuminating illustration of the effects of the estimation bias and how to overcome it, one might argue that this single spin example constitutes a highly specific edge case, which might not be relevant for general many-body systems. 
We now expand on the previous example by considering the following interacting system
\begin{equation}
   H = -J\sum_{i=1}^{N-1} \hat{\sigma}^z_i \hat{\sigma}^z_{i+1} + g \sum_{i=1}^N \hat{\sigma}^y_i. \label{eq:bias_interaction}
\end{equation}
with $L=10$ spins and coupling $J/g=0.1$.
Starting again from an initial product-state polarized in the positive x-direction, we evolve the system using a Restricted Boltzmann Machine (RBM) until a time $g\, t_\mathrm{max} = 2$. % for different numbers of samples and cutoff parameters $\epsilon$.
Fig.~\ref{fig:bias_cutoff} b) shows the time evolution of the x-magnetization for $N_\mathrm{S}=1000$ samples and different values of the cutoff. 
The largest deviation from the exact reference time evolution is observed for $\epsilon=0$, corresponding to the Born-distribution. 
The inset shows a more quantitative analysis, by plotting the infidelity $1-\mathcal{F}=1-|\braket{\psi_\mathrm{exact}|\psi_\mathrm{RBM}}|^2$ with respect to the exact result at $tJ=2$ as a function of the number of samples used. 
Choosing any non-vanishing $\epsilon$ reduces the infidelity by up to two orders of magnitude compared to sampling based on the Born distribution, with the optimal value lying at $\epsilon \approx 10^{-3}$. 
As a consequence, using a finite cutoff $\epsilon$ allows to reduce the number of samples used for the time evolution, without compromising on its quality.

Having established the general usefulness of the cutoff-based importance sampling scheme, we turn towards a final, non-trivial example of a quench in the one dimensional transverse field Ising model (TFIM), with the following Hamiltonian
\begin{equation}
H = -J \sum_{i=1}^{N-1}  \hat{\sigma}_i^z \hat{\sigma}_{i+1}^z + g \sum_{j=i}^{N} \hat{\sigma}_i^x \label{eq:tfim}
\end{equation}
The dynamics of this model, in particular for quenches to the critical point $g/J=1$, is known to be particularly difficult to capture for neural quantum states~\cite{czischek_2018}, which in the following we investigate under the point of view of sampling.
Starting from an x-polarized state $\ket{\psi_0}=\frac{1}{2^N}\sum_{\vec s}\ket{\vec s}$, we perform the time evolution after a quench to the critical point on a chain-lattice of size $L=20$. 
We are using a ResNet architecture~\cite{liu_2025resnet} with two blocks and a total of $N_\mathrm{params}=12200$ parameters, taking $N_\mathrm{samples}=4\cdot10^4$ samples in each step.
\begin{figure}[h!]
    \includegraphics[width=0.9\columnwidth]{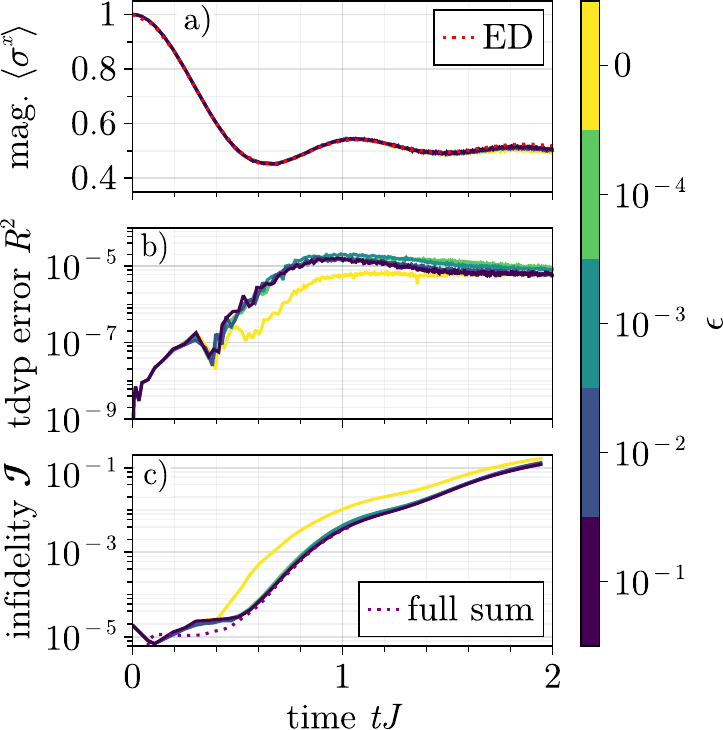}
    \caption{Time evolution of an initially x-polarized state after a quench to the critical Ising model for a system size $L=20$ using different cutoffs and $4\cdot10^4$ samples. (a) The time evolution of the magnetization matches the exact reference results for all cutoffs up to a time $tJ=1.3$, with slight deviations appearing between different cutoffs at around $tJ=1.6$. (b) shows the TDVP error $R^2$, a measure for the projection error at each time step. (c) Time evolution of the infidelity between the variational and reference state. The exact solution is followed with higher precision for all cases of $\epsilon>0$ and the infidelities remain very close to those obtained by performing the full summation in the TDVP equation.} \label{fig:tfim_cutoff}
\end{figure}

The time evolution of the magnetization, shown in Fig.~\ref{fig:tfim_cutoff}~a), matches the reference results obtained using the Lanczos method for all cutoffs up to a time of around $tJ=1.3$. Differences between different values of the cutoff become apparent starting from $tJ=1.6$, with curves corresponding to non-zero $\epsilon$ lying closer to the reference result.
The TDVP error $R^2$, introduced in~\cite{schmitt2020quantum} and used as a measure for the local projection error, appears to be smaller when sampling using the Born distribution compared to the cutoff case, see Fig.~\ref{fig:tfim_cutoff}~b). 
However, we want to note, that this might potentially be misleading -- the error is calculated during the TDVP step with sampled quantities affected by the estimation bias, potentially leading to an underestimation of the error. 
A better indicator for the quality of the time evolution is the infidelity between the variational and exact state at each timestep, shown in Fig.~\ref{fig:tfim_cutoff}~c). We observe that the infidelity starts growing noticeably later in time for curves corresponding to nonzero $\epsilon$ compared to $\epsilon=0$, without any discernible differences between the different non-zero cutoff values. Most notably, we observe that sampling using non-zero cutoff closely reproduces the result obtained when performing the summation over the full Hilbertspace.

\begin{figure}[h!]
    \includegraphics[width=\columnwidth]{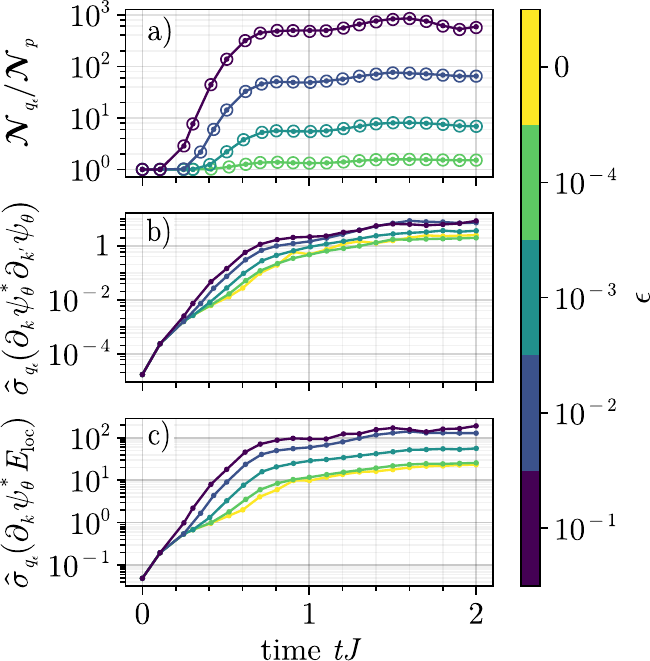}
    \caption{(a) Time evolution of the ratio of normalization constants~\eqref{eq:norm_ratio_estimation} for different choices of the cutoff. Dots denote values obtained through sampling using $N_S=4\cdot10^4$, circles denote results obtained through full summation, showing an excellent agreement between the two approaches. (b) and (c) show the time evolution of the mean over the component-wise variance of the first two terms appearing in~\eqref{eq:traditional_vmc_deriv}, respectively, for different choices of the $\epsilon$. Starting from $tJ=0.5$, a larger cutoff parameter leads to an increase in the estimation variance; the overall increase, however, remains limited---especially for the smallest values of $\epsilon$.}  \label{fig:tfim_snr}
\end{figure}

Fig.~\ref{fig:tfim_snr}~a) showcases the evolution of the ratio of normalization constants, comparing an exact calculation to the sampled estimation~\eqref{eq:norm_ratio_estimation}, denoted by circles and dots respectively. 
Expectedly, this ratio departs from $\frac{\mathcal{N}_{q_\epsilon}}{\mathcal{N}_p}=1$ when the wave function amplitudes starts exhibiting zero crossing, at around $tJ\approx0.5$.
We want to note that we are considering five different sampling realizations, plotting both the mean and the standard deviations as error bars, with the latter being so small it cannot be seen on the scale of the plot, confirming that a reliable estimation of the ratio of normalization is possible for the chosen class of distributions, even at larger cutoff values.

Finally, having resolved the estimation bias issue, we want to shortly comment on the standard deviation of the introduced importance sampling estimator.
In particular, we are interested in the mean over the component-wise standard deviation, defined by $\hat{\sigma}_q$ in Eq. \eqref{eq:snis_variance}, of the first and second quantity in Eq.~\eqref{eq:traditional_vmc_deriv}, plotted as functions of time in Fig.~\ref{fig:tfim_snr}~b) and c) respectively
\footnote{An alternative probe of fluctuations that is considered in recent work is the mean component-wise signal-to-noise-ratio (SNR) \cite{misery2025}. We, however, resort to the plain standard deviation, because vanishing SNR of zero components is not necessarily informative about the estimation cost.}.

While $\hat{\sigma}_{q_\epsilon}\Big(\partial_k\psi_\theta^*\, \partial_{k'}\psi_\theta\Big)$ does not strongly depend on $\epsilon$, we observe that $\hat{\sigma}_{q_\epsilon}\Big(\partial_k\psi_\theta^*\, E_\mathrm{loc}\Big)$ is smaller for smaller $\epsilon$. 
This behavior has two potential reasons: on one hand, the quantity to be estimated contains $E_\mathrm{loc}$, which itself is proportional to the wave function amplitude $\psi_\theta$. Because of that, samples with respect to $|\psi_\theta(\vec s)|$ cover the region with the most dominant contributions of the estimator, leading to a lower standard deviation.
On the other hand, since the standard deviations are being sampled themselves, they are subject to the estimation bias as well, leading to a potential underestimation of the fluctuation, especially for small $\epsilon$.

However, we want to point out that the observed differences in the standard deviations are at most one order of magnitude.
Specifically for $\epsilon=10^{-4}$, which is already sufficient for eliminating the estimation bias, we observe no significant increase in the estimator standard deviation.

In the next section we shift the paradigm and explore how TCI can be used to circumvent sampling in the traditional MC sense on the example of the TFIM.

\subsection{Tensor Cross Interpolation}

Similar to the end of the last section, we are considering quenches to the critical point of the TFIM~\eqref{eq:tfim}, starting from a state polarized in the x-direction.

The analysis has been structured as follows: having performed a time evolution of the quench for different system sizes using an NQS and the traditional t-VMC algorithm, we evaluate the QGT and force vector using the algorithm based on TCI as proposed in~\ref{sec:TCI_intro} for different maximal bond dimensions at various points in time. 
In order to gauge how well the TCI representation is able to encode these quantities, we compare them to their exact counterparts obtained through full, i.e. exact summation by considering the relative error based on the Frobenius norm
\begin{equation}
    \delta_X = \frac{||X-X_\mathrm{exact}||_2}{||X_\mathrm{exact}||_2}
\end{equation}
with $X=\hat{\mathcal{S}}/\hat{\mathcal{F}}$, for the QGT and force vector, respectively.
Since these objects govern the TDVP-equation, their accuracy can be used to directly gauge the accuracy of the scheme.

\begin{figure}[h!]
    \includegraphics[width=\columnwidth]{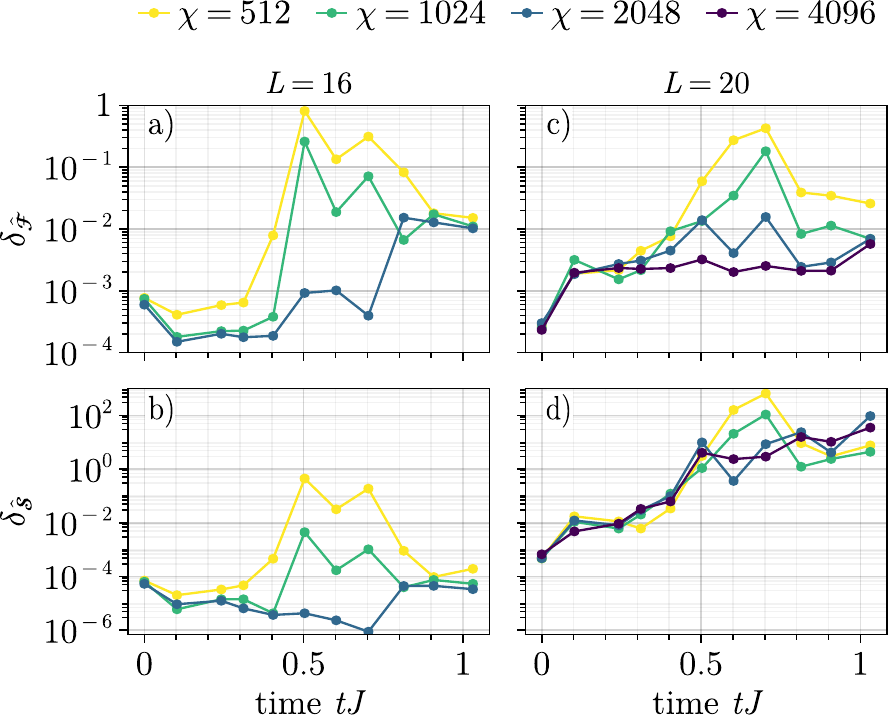}
    \caption{Investigation of the relative error of the force vector (a and c) and QGT (b and d) obtained for different system sizes using TCI with various maximal bond dimensions as functions of time. Performing the same analysis using MC-sampling and $N_S=4\cdot10^4$ samples yielded relative errors of the order of $\delta\approx10^{-2}$, which in the following we use as a reference benchmark result. At $L=16$ (a and b), increasing the bond dimension successfully suppresses the relative error below said reference benchmark. At $L=20$ (c and d) estimating the QGT even at maximal bond dimension of $\chi=4096$ becomes harder, leading to large relative errors, signaling the limitations and failure of the TCI-based approach.} \label{fig:tci_err}
\end{figure}

These comparisons are shown in Fig.~\ref{fig:tci_err} a) and b) for a system size of $L=16$ and c) and d) for $L=20$. The initial time evolution is performed using a convolutional neural network (CNN) with a filter size of $f=10$ and $N_\mathrm{channels}=20$/$40$ for $L=16$/$20$, respectively.
For the TCI algorithm itself, we set a tolerance of $\epsilon_\mathrm{tci}=10^{-4}$ as well as various bond dimensions ranging from $\chi_{\max}=512$ to $\chi_{\max}=4096$. 

Note that throughout our testing, calculating the same error quantities via MC-sampling resulted in an error of  $\delta_{\hat{\mathcal{S}}/\hat{\mathcal{F}}}\approx 10^{-2}$, which we use in the following as a guide to gauge the accuracy of the TCI algorithm.
Starting with the smaller lattice $L=16$, we observe that both the force vector and QGT, Fig.~\ref{fig:tci_err}~a) and b) respectively, can be estimated reliably for the largest considered bond dimension of $\chi_{\max}=2048$. Expectedly, increasing said bond dimension improves the accuracy of the results. 
Note that due to the chosen structure of physical indices and the resulting tensor networks (cf.~Fig.~\ref{fig:mps_eps_gamma}), the maximal bond dimension of the networks itself is capped at $\chi_{\max}^{\mathrm{TCI}_\epsilon} = 256$ for the local energy term and $\chi_{\max}^{\mathrm{TCI}_\gamma} = 3200$ for the gradient term.

For the larger lattice $L=20$ however, the estimation becomes significantly more difficult, as these values rise to $\chi_{\max}^{\mathrm{TCI}_\epsilon} = 1024$ for the local energy term and $\chi_{\max}^{\mathrm{TCI}_\gamma} = 16384$ for the gradient term.
While the estimation of the $\hat{\mathcal{F}}$-vector Fig.~\ref{fig:tci_err}~c) is still feasible using $\chi_{\max}=4096$, with the resulting deviations lying between $10^{-4}< \delta_{\hat{\mathcal{F}}}<10^{-2}$, the estimation of the $\hat{\mathcal{S}}$-matrix Fig.~\ref{fig:tci_err}~d) fails, with errors quickly climbing up to $\delta_{\hat{\mathcal{S}}} \approx 10$. 

Increasing the bond dimension even further quickly becomes computationally infeasible, with the estimation of a single $S$-matrix and $F$-vector taking around $t_\mathrm{sim}\approx20h$ at $\chi_{\max}=4096$.
While the proposed approach based on actively learning the most relevant samples constitutes an interesting alternative view to the MC-sampling paradigm, in its current formulation, the computational cost becomes prohibitive already at moderate system sizes, which can be traced back to the lack of a low-rank structure in the gradient of the ansatz with respect to its variational parameters.

\section{Discussion}
The bias issue is a fundamental problem inherent to traditional VMC approaches. 
In this work, we investigate different ways to circumvent it, either by modifying the sampling distribution via self-normalized importance sampling, or attempting to actively choose the relevant samples for all quantities relevant to the VMC algorithm via TCI. 

For self-normalized importance sampling, care needs to be taken in order to not introduce a different, artificial bias and not to increase the standard deviation of the estimator. 
We confirm that our proposed, cutoff-based deformation of the Born distribution not only solves the bias issue, providing time evolution results on par with what is obtained using full summation, but also does not lead to an increased estimation variance.
Thereby, our approach constitutes a well-defined variant of t-VMC, that is obtained by minimal modifications of the original formulation of the algorithm and straightforward to implement in practice.

In addition, we have explored an alternative approach to VMC, leveraging the TCI to obtain efficiently contractable tensor network representations of all quantities in the TDVP equation.
While this provides a systematic control over the quality of the estimation via the tunable bond dimension of the tensor network representation, its main drawback lies in the necessity of a low-rank representation of the gradient of the wavefunction with respect to the variational parameters. 
Additionally, as opposed to Monte Carlo sampling which is highly parallelizable, the inherent sequentiality of the TCI algorithm leads to an overall increased runtime at the same level of accuracy.
Nonetheless, there are numerous avenues left to be explored with this approach. First of all, the network architecture proposed in this work was chosen due it's simple implementation, however it introduces a strong asymmetry between the spin- and parameter degrees of freedom. 
By correctly disentangling the variational parameters or changing the architecture to one respecting that asymmetry, e.g. a tree-like structure, one could potentially significantly reduce the bond dimension. 
Furthermore, in this work we opted to learn representations of the gradient of the wave function and the local energy, used to subsequently build the other quantities necessary for each TDVP step. 
As an alternative approach to incorporate TCI into variational Monte Carlo, one can shift the point of view by e.g. directly learning low rank tensor-network representations of the quantum geometric tensor and force vector. While this would not get rid of the sampling bias issue, this could provide avenues for finding new, optimized update rules.

\begin{acknowledgments}
WK and MS were supported through the Helmholtz Initiative and Networking Fund, Grant No. VH-NG-1711.
\end{acknowledgments}

\appendix
\section{Explicit form of importance-sampled quantities} \label{app:derivation_tdvp}

In this section we summarize the explicit form of all importance sampled quantities that are relevant to the TDVP algorithm, i.e. ones that appear in Eqs.~\eqref{eq:Q} and \eqref{eq:F}. 
This is done similarly to~\eqref{eq:F_first_term_impsamp}, where introduce artificial $1$'s inform of $\frac{q(\vec s)}{q(\vec s)}$, leading to

\begin{align}
    \frac{\braket{\partial_k\psi_\theta|\hat H|\psi_\theta}}{\langle\psi_\theta|\psi_\theta\rangle}
    &=
    \mathbb{E}_{Q}
    \Bigg[\frac{\mathcal{N}_q}{\mathcal{N}_p}\frac{\partial_k\psi_\theta^*\, E_\mathrm{loc}}{q}\Bigg]\, ,
    \nonumber\\
    \frac{\braket{\partial_k\psi_\theta|\partial_{k'}\psi_\theta}}{\langle\psi_\theta|\psi_\theta\rangle}
    &=
    \mathbb{E}_{Q}
    \Bigg[\frac{\mathcal{N}_q}{\mathcal{N}_p}\frac{\partial_k\psi_\theta^*\, \partial_{k'}\psi_\theta}{q}\Bigg]\, ,
    \nonumber\\
    \frac{\braket{\psi_\theta|\hat H|\psi_\theta}}{\langle\psi_\theta|\psi_\theta\rangle}
    &=
    \mathbb{E}_{Q}
    \Bigg[\frac{\mathcal{N}_q}{\mathcal{N}_p}\frac{\psi_\theta^*\, E_{\text{loc}}}{q}\Bigg]\, , \nonumber \\
    \frac{\braket{\psi_\theta|\partial_{k'}\psi_\theta}}{\langle\psi_\theta|\psi_\theta\rangle}
    &=
    \mathbb{E}_{Q}
    \Bigg[\frac{\mathcal{N}_q}{\mathcal{N}_p}\frac{\psi_\theta^*\, \partial_{k'}\psi_\theta}{q}\Bigg]\, .
    \label{eq:traditional_vmc_deriv}
\end{align}
As before, the ratio of normalization constants $\frac{\mathcal{N}_q}{\mathcal{N}_p}$ is being estimated via Eq.~\eqref{eq:norm_ratio_estimation}.

\bibliography{refs}

\end{document}